\documentclass[fleqn,usenatbib]{mnras}

\usepackage[T1]{fontenc}
\usepackage{ae,aecompl}

\usepackage{graphicx}	
\usepackage{amsmath}	
\usepackage{amssymb}	
\usepackage{lscape}

\title[Asteroseismology of PG~1541$+$651 and BPM 31594]{Asteroseismology of PG 1541$+$651 and BPM 31594 with TESS.}

\author[Romero et al.]{Alejandra D. Romero$^{1}$\thanks{E-mail: alejandra.romero@ufrgs.br},  Gabriela Oliveira da Rosa$^1$, S. O. Kepler$^{1}$,  Paul A. Bradley$^{2}$, 
\newauthor Murat Uzundag$^{3,4}$, Keaton J. Bell$^{5,6}$, J. J. Hermes$^{7}$,  and G. R. Lauffer$^{1}$ \\
$^{1}$Instituto de F\'\i sica, Universidade Federal do Rio Grande do Sul, Av. Bento Gon\c{c}alves 9500, Brazil\\
$^{2}$XCP-6, MS F-699 Los Alamos National Laboratory, Los Alamos, NM 87545\\
$^{3}$Instituto de F\'{\i}sica y Astronom\'{\i}a, Universidad de Valpara\'{\i}so, Gran Breta\~na 1111, Playa Ancha, Valpara\'{\i}so 2360102, Chile\\
$^{4}$European Southern Observatory, Alonso de Cordova 3107, Santiago, Chile\\
$^{5}$DIRAC Institute, Department of Astronomy, University of Washington, Seattle, WA-98195, USA\\
$^{6}$NSF Astronomy and Astrophysics Postdoctoral Fellow\\
$^{7}$Department of Astronomy \& Institute for Astrophysical Research, Boston University, 725 Commonwealth Ave., Boston, MA 02215, USA\\
 \\
}

\date{Accepted XXX. Received YYY; in original form ZZZ}

\pubyear{2018}

\begin{document}
\label{firstpage}
\pagerange{\pageref{firstpage}--\pageref{lastpage}}
\maketitle

\begin{abstract}
We present the photometric data from TESS for two known ZZ Ceti stars, PG~1541+651 and BPM~31594. Before TESS, both objects only had observations from short runs from ground-based facilities, with three and one period detected, respectively. The TESS data allowed the detection of multiple periodicities, 12 for PG~1541$+$651, and six for BPM~31594, which enables us to perform a detailed asteroseismological study. 
For both objects we found a representative asteroseismic model with canonical stellar mass $\sim 0.61 M_{\odot}$ and thick hydrogen envelopes, thicker than $10^{-5.3} M_*$. The detection of triplets in the Fourier transform also allowed us to estimate mean rotation periods, being $\sim$ 22 h for PG 1541+651 and 11.6 h for BPM~31594, which is consistent with range of values reported for other ZZ Ceti stars.

\end{abstract}




\section{Introduction}
\label{sec1}

White dwarf stars are the endpoint of the evolution for all stars with initial masses up to 8.5-12 $M_{\odot}$ \citep{2007A&A...476..893S,2015MNRAS.446.2599D,2018MNRAS.480.1547L} when we consider single stellar evolution. This implies that more than 97\% of the stars in the Milky Way, including the Sun, will end their lives as white dwarf stars. Therefore, the white dwarf population conveys an important
record of the evolution of all stellar populations in our Galaxy. The most numerous class of white dwarf stars are the hydrogen atmosphere DAs, comprising $\sim$ 87\% of all white dwarfs known to date \citep{2019MNRAS.486.2169K,2021MNRAS.507.4646K}. In addition, variable DA white dwarf stars are also the most numerous class of pulsating white dwarfs. They are known as ZZ Ceti stars or DAVs, comprising $\sim$ 80\% of all known pulsating white dwarfs \citep[e.g.][]{2021A&A...651A..14B}. ZZ Ceti stars show periodic brightness variations due to gravity driven g-mode pulsations, 
with periods from 70 to 3000~s and variation amplitudes of 1–60 ppt \citep[e.g.][]{2008PASP..120.1043F,2008ARA&A..46..157W,2017EPJWC.15201011K,2019A&ARv..27....7C}. The excitation mechanism is related to the increase in opacity at the base of the hydrogen envelope, due to partial ionization of hydrogen \citep{1981A&A...102..375D,1982ApJ...252L..65W} and later the ability of the convection zone to react quickly to changes in thermal structure \citep{1991MNRAS...251..673B,1999ApJ...511..904B}. This occurs
in a narrow range of effective temperature, between $13\, 500$ K and $10\, 500$ K, depending on the stellar mass. The instability strip for ZZ Cetis is pure, meaning that all DA white dwarfs should show photometric variability within these effective temperature range \citep{2007A&A...462..989C, 2011ApJ...743..138G}.

As we move through the ZZ Ceti instability strip, there is a change in the behaviour of the pulsation spectrum  \citep{1993BaltA...2..407C, 2006ApJ...640..956M}. The hot ZZ Cetis, near the blue edge of the instability strip, show stable sinusoidal or sawtooth light curves, with a few short periods (<350 s) and small variation amplitudes (1.5–20~mma). On the red edge, the cool ZZ Cetis show non-sinusoidal light-curves and a collection of long periods (> 650 s), with large  amplitude variations (40–110~mma) that suffer from severe mode interference, leading to the presence of linear combination frequencies and/or harmonics in the Fourier transform. Around $11\, 500$ K, the warm-like ZZ Cetis show mixed characteristics from hot and cool members for short and long periods, respectively. 

Asteroseismology applied to white dwarf stars allows us to study the inner structure and the evolutionary properties of these objects through the observed period spectrum, in the same way as quakes on Earth allow for the study of the inner core. The technique consists of a comparison between the observed periods and the theoretical periods computed from representative models, as inversion is not doable at present. Each pulsation mode propagates in a specific region, providing information on that particular place inside the star, where its amplitude has a maximum weight. In particular, the thickness of the hydrogen and helium layers \citep{2008MNRAS.385..430C,2009MNRAS.396.1709C,2012MNRAS.420.1462R,2013ApJ...779...58R}, the chemical composition of the inner core and an estimate of the $^{12}$C$(\alpha,\gamma)$O$^{16}$ reaction rate \citep{2002ApJ...573..803M,2017A&A...599A..21D,2022MNRAS.513.1499P,2022ApJ...935...21C}, crystallization \citep{1999ApJ...526..976M, 2013ApJ...779...58R}, rotation velocity \citep{2017ApJ...841L...2H}, and the properties of the convective regions \citep{2007ASPC..372..635M,2020ApJ...890...11M}, can be determined from the observed period spectrum through asteroseismology. In addition, the determination of the rate of period change \citep{2005ApJ...634.1311K} can be used to study elementary particles such as neutrinos \citep{2004ApJ...602L.109W} and axions \citep{2012MNRAS.424.2792C, 2013ApJ...771...17M, 2016JCAP...07..036C,2021ApJ...906....7K}. 

Since the discovery of the first ZZ Ceti star, HL Tau 76
\citep{1968ApJ...153..151L}, there are more than 400 ZZ Cetis reported to date \citep[see for instance][]{2016IBVS.6184....1B,2019A&ARv..27....7C,2019MNRAS.490.1803R,2020AJ....160..252V,2021ApJ...912..125G, 2022MNRAS.511.1574R}. However, for most ZZ Cetis, only a limited number of observed periods are known, usually detected as a result of short runs from the discovery paper, and no follow-up observations were performed. 

The increase in the number of new ZZ Cetis and new detected periods was boosted by recent space-based observations. For instance, the TESS satellite \citep{2014SPIE.9143E..20R,2015JATIS...1a4003R}, launched in 2018 April 18, observed several ZZ Ceti candidates and known pulsators \citep[see for example][]{Bognar2020, 2022MNRAS.511.1574R}. This satellite observes between 150
and 300 white dwarf stars every month. Each observation run last for a total of 27 days, in a 120 s-–cadence for selected objects, and a 20 s--cadence mode for a limited sample of bright objects. 

In particular, PG~1541+651 and BPM~31594 are two warm-like ZZ Cetis with three and one periods, respectively, detected from ground-based discovery observations \citep[][]{2000A&A...355..291V,1976ApJ...210L..35M}. The number of detected modes considerably increased after they were observed by the TESS satellite, allowing a more detailed asteroseismological analysis. 
In this work, we present observations from TESS for PG~1541+651 and BPM~31594, along with a detailed asteroseismological analysis. The paper is organized as follows: in section \ref{sec2} we present the results from previous observation runs for both objects. We describe the data reduction process and results from TESS observations in section \ref{sec3}. Section \ref{sec4} is devoted to a detailed asteroseismological analysis of the targets, and we summarized our findings in section \ref{conclusion}.


\section{selected targets}
\label{sec2}

\cite{2000A&A...355..291V} reported the detection of photometric variability in PG~1541$+$651 (TIC~458484139), based on $\sim$9~h observations performed on the 2-m 
Pic du Midi and the 1.93-m Haute Provence Observatory telescopes. Three periods were identified at 689, 564 and 757~s, with the 689~s period having the largest amplitude\footnote{PG~1541+651 can also be found in the literature as PG~1541+650.}. These periods show typical values for a ZZ Ceti in the middle of the instability strip \citep{2006ApJ...640..956M}.
The spectroscopic effective temperature and surface gravity were determined by \cite{2011ApJ...743..138G}, $T_{\rm eff} = 11\,560\pm 250$~K and $\log g = 8.12\pm 0.038$, after applying the 3D convection correction \citep{2013A&A...559A.104T}, leading to a stellar mass of $0.67\pm 0.03 M_{\odot}$. Later on, \citet{2021MNRAS.508.3877G}, using photometry and parallax from Gaia eDR3, found  $T_{\rm eff} = 11\,607\pm 203$~K and $\log g = 8.035\pm 0.028$, with a significantly lower stellar mass of $0.626\pm 0.018 M_{\odot}$. 

The pulsational variability of BPM~31594 (TIC~101014997) was first reported by \citet{1976ApJ...210L..35M} based on $\sim$13~h of observations at the Sutherland observing station of the South African Astronomical Observatory, using the 0.76-m and 1.02-m telescopes. He found two periods at 617~s and 314~s, with an amplitude of 0.18~mag. \citet{1992MNRAS.258..415O}, based on $\sim$300 h run at the 0.76- 1.02- and 1.9-m reflectors at the Sutherland observing station of the South African Astronomical Observatory, also reported the detection of a period of 617~s, along with several harmonics.

\cite{2011ApJ...743..138G} obtained the atmospheric parameter for BPM~31594, being  $T_{\rm eff} = 11\,500 \pm 250$~K and $\log g = 8.05\pm 0.038$ after applying the 3D convection correction. With these values, the stellar mass is $0.63\pm 0.03 M_{\odot}$. Later \citet{2017PhDT........20F} found a slightly higher effective temperature of $11\,786 \pm 22$~K and a lower surface gravity $\log g = 8.029\pm 0.006$, corresponding to a stellar mass of $0.615\pm 0.005 M_{\odot}$, similar to that found by  \cite{2011ApJ...743..138G}. 

Note that, as is the case for many ZZ Ceti stars, no further follow-up observations were published for these objects until the TESS satellite.

\section{Data analysis}
\label{sec3}

BPM~31594 was observed in Sectors 3 and 4 with 120~s--cadence, and in Sectors 30 and 31 with both 120~s and 20~s cadence. PG~1541$+$651 was observed in Sectors 14 to 17 and 21 to 24 with 120~s cadence and in Sectors 41, 47, 48, 50, and 51 in 120~s and 20~s cadence. 

We downloaded all light curves for BPM~31594 and PG~1541$+$651 from the Mikulski Archive for Space Telescopes, which is hosted by the 
Space Telescope Science Institute (STScI)\footnote{http://archive.stsci.edu/} in FITS format. The data were processed based on the Pre-Search Data Conditioning pipeline \citep{2016SPIE.9913E..3EJ}.
We extracted times and fluxes (PDCSAP FLUX) from the FITS files. The times are given in barycentric corrected dynamical Julian days \citep[BJD – 2457000, corrected for leap seconds, see][]{2010PASP..122..935E}. 
For each sector, the fluxes were converted into fractional variations from the mean, that is, differential flux $\Delta I/I$, and transformed into amplitudes in parts-per-thousand (ppt). The ppt unit corresponds to the milli-modulation amplitude (mma) unit\footnote{1 mma= 1/1.086~mmag= 0.1\% = 1 ppt; see, e.g., \citet{2016IBVS.6184....1B}.}.  The contamination level from other stars in the image is low, with CROWDSAP\footnote{The CROWDSAP level indicates the ratio of the target flux to the total flux.} of 0.94 and 0.704 for BPM~31594 and PG~1541$+$651, respectively. We sigma-clipped the data at 5$\sigma$ to remove the outliers that appear above five times the median absolute deviation, that is, that depart from the median by 5$\sigma$. 

We computed the Fourier transforms (FTs) for all light curves and looked for pulsations signatures above the detection limit, defined as 5$\left<A\right>$.
For pre-whitening, we employed our customized tool, in which, using a nonlinear least-squares method, we simultaneously fit each pulsation frequency in a waveform $A_i\sin(\omega_1 t + \phi)$, with $\omega = 2\pi / P$, and $P$ the period. This iterative process was run, starting with the highest peak, until no peak appeared above the detection limit. The uncertainties in the frequencies are estimated as $1/T$, where $T$ is the duration of each block, and vary from 0.11~$\mu$Hz to 0.43~$\mu$Hz.

\subsection{PG~1541+651}

\begin{figure*}
	\includegraphics[width=\textwidth]{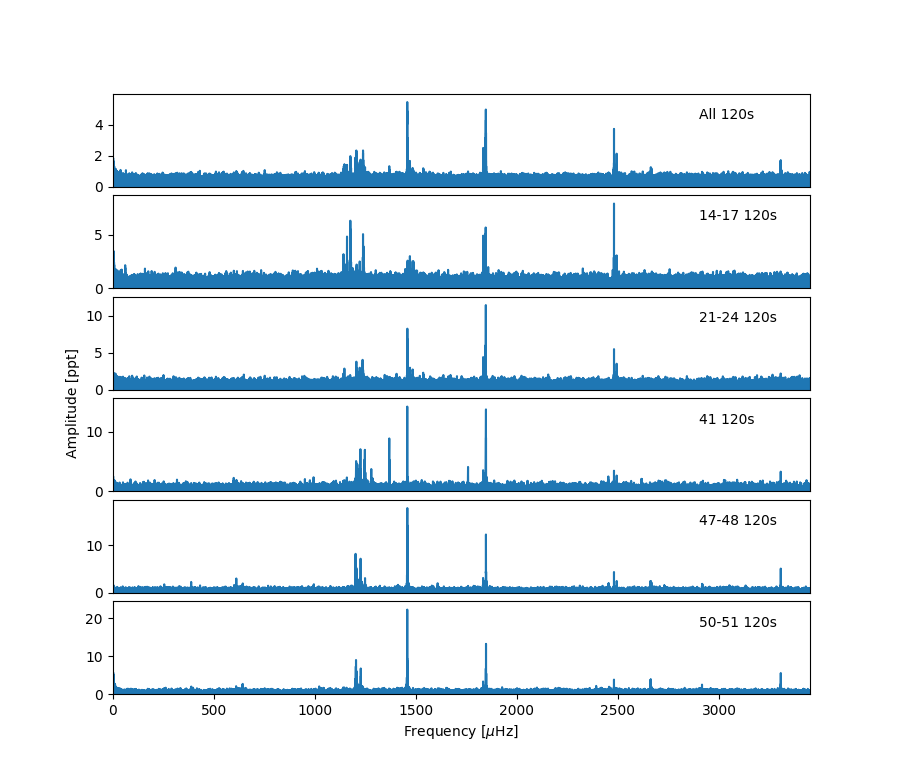}
    \caption{Fourier transform for the data for PG~1541+651. The top panel corresponds to the FT of all concatenated sectors. The data from second to sixth panels, from top to bottom, correspond to the blocks including Sectors 14--17, 21--24, 41, 47--48 and 50--51, with 120s--cadence. Note that the amplitude scale is different for each plot.}
    \label{FT-PG}
\end{figure*}

The ZZ Ceti star PG~1541+651 was observed by TESS in Sectors 14--17, 21--24, 41, 47--48, and 50--51. For Sectors 41, 47, 48, 50, and 51 the observations were taken with 20s--cadence, while for the previous sectors, the data correspond only to 120s--cadence. 
For this object, we separate the data into five blocks, corresponding to continuous observations runs, to avoid possible spurious signals due to long gaps in the data. For each block, we obtained the Fourier transform (FT) and identify the characteristic frequencies. Figure \ref{FT-PG} shows the FT for all concatenated sectors, Sectors 14--17, Sectors 21--24, Sector 41, Sectors 47--48 and Sectors 50--51, from top to bottom. Note that the amplitude scale is not the same for all the plots. The data corresponds to 120s--cadence observations for all blocks, as there are no periods detected below the Nyquist limit, corresponding to a period of 240~s.

The frequencies detected from each block are listed in Table \ref{table-multiplets}, along with the corresponding periods and amplitudes. As can be seen from this table, there are only two frequencies that are present in all five blocks, but most of them are detected in four of the five blocks. Other frequencies are present in some blocks, but not detected in others. Note that not all the frequencies correspond to independent pulsation modes, since there are several multiplets in the period spectra (see section \ref{triplets}).

As can be seen from Figure~\ref{FT-PG}, and also from the values listed in Table~\ref{table-PG-obs}, the amplitudes of the peaks in the FT vary from block to block. For the first block (14-17) the dominant mode, that with the highest amplitude in the FT, is at a frequency of 2480.5871~$\mu$Hz (403.1304~s). In the second block (21-24) the amplitude of that mode decreases considerably, while the mode with a frequency of $\sim$1847~$\mu$Hz ($\sim$ 541~s) shows the highest amplitude in the FT. Finally, the amplitude for the mode with $\sim$1847 $\mu$Hz decreases from the third to the fifth block, making the mode with a frequency of $\sim$ 1458~$\mu$Hz ($\sim$685~s) the dominant mode from Sectors 41 to 51. This change in amplitude, from a short to a longer period in a scale of a few years is not likely to be related to convection \citep{2020ApJ...890...11M}. 

In general, peaks corresponding to high-frequencies, higher than $\sim$1250~$\mu$Hz are narrow, with widths roughly matching the spectral window of the observations.   
The peaks with frequencies below 1248\,$\mu$Hz (periods longer than $\sim$ 800~s) show a complex structure in the FT for all the blocks, spreading their power over a broad band in the FT. This behaviour is in agreement with the dichotomy found by  \citet{2017ApJS..232...23H}, based on 27 DAVs observed with the {\it Kepler} satellite. These authors found that peaks corresponding to periods longer than $\sim$800~s have substantially broader mode widths than those with periods shorter than $\sim$800~s, and are most likely representative of phase--unstable single modes, reminiscent of a damped harmonic oscillator. Later, \citet{2020ApJ...890...11M} showed that the depth of the convection zone changes during the pulsation cycle, causing the reflection of the outgoing travelling wave to not be coherent. Since in most cases, modes with periods longer than $\sim$800~s propagate all the way to the base of the surface convection zone, these will be affected by the time-dependent position of the outer turning point. 

\begin{table*}
\centering
\caption{List of periods and amplitudes for PG 1541+651 for each block (top row).}
\label{table-PG-obs}

\begin{tabular}{cccccccccccc}
\hline\hline
14-17 &   &  21-24  &   &   41   &   & 47-48 &  & 50-51 &   & all &  \\ 
\hline
$\Pi$ & Amp & $\Pi$ & Amp  &  $\Pi$ & Amp &  $\Pi$ & Amp & $\Pi$ & Amp & $\Pi$ & Amp \\

 s & ppt & s & ppt & s & ppt & s & ppt & s & ppt & s & ppt\\
\hline
$\cdots$ & $\cdots$ & $\cdots$  & $\cdots$ & 302.5562 & 3.3489 &  302.5351 & 5.8632 &  302.5338 & 8.2818 & 302.5344 & 4.7417 \\
$\cdots$ & $\cdots$ & $\cdots$  & $\cdots$ & $\cdots$ & $\cdots$ & 375.7850 & 4.2753 & 375.1835 & 5.2731 & 375.5697 & 2.8014 \\
401.0212  & 1.5348 &  401.0175  & 3.3980  &  $\cdots$  &  $\cdots$ & 400.9785 & 2.9746 &  $\cdots$ & $\cdots$ & 400.9768 & 2.6916 \\
403.1304  & 7.8849 &  403.1295  & 5.1636  &  403.1435 & 3.4118 &  403.1490 & 4.6771 &  403.1557 & 5.2909 & 403.1195 & 4.2203\\
541.6983  & 7.5621 &  541.5875  & 13.2365 &  541.3547 & 13.7774 & 541.3733 & 14.2022 & 541.2986 & 17.9046 & 541.2837 & 8.2155\\
545.4038  & 4.6067 &  545.4320  & 5.1746  & 545.5416 & 3.2471 &  545.5766 & 3.3115 &  $\cdots$ & $\cdots$ & 545.5728 & 3.1326 \\
$\cdots$ & $\cdots$ & $\cdots$ & $\cdots$ &  568.6268 & 3.9720 & $\cdots$ & $\cdots$  &  $\cdots$ & $\cdots$ & $\cdots$ & $\cdots$ \\
$\cdots$ & $\cdots$ & $\cdots$  & $\cdots$ & $\cdots$ & $\cdots$ & 621.8078 & 2.4711  &  $\cdots$ & $\cdots$ & $\cdots$ & $\cdots$\\
$\cdots$ & $\cdots$ & 679.5963  & 3.0513  &  $\cdots$ & $\cdots$ & $\cdots$ & $\cdots$ &  $\cdots$ & $\cdots$ & $\cdots$ & $\cdots$\\
$\cdots$  &  $\cdots$ & 685.1075  & 7.4502  & 685.8707 & 3.9720 &  685.7314 & 18.3264 & 685.8732 & 27.8657 & 685.9228 & 9.3018 \\ 
$\cdots$  &  $\cdots$ & $\cdots$ & $\cdots$  & 730.4653 & 8.8848 & $\cdots$ &  $\cdots$ &  $\cdots$ & $\cdots$ & 730.4495 & 2.8126 \\
$\cdots$ &  $\cdots$ & $\cdots$  &  $\cdots$ & 781.2597 & 3.5981 & $\cdots$ & $\cdots$  &  $\cdots$ & $\cdots$ & $\cdots$ & $\cdots$\\
$\cdots$ &  $\cdots$ & $\cdots$  & $\cdots$  & 801.2480 & 6.8186 & 800.5709 & 3.0613 & $\cdots$ & $\cdots$ & 801.0640 & 2.9695 \\   
806.5962  & 6.2639 &  808.3629  & 5.2433  & $\cdots$ &  $\cdots$ & 810.6814 & 2.3961 & $\cdots$ & $\cdots$ & $\cdots$ & $\cdots$ \\
$\cdots$ &  $\cdots$ & $\cdots$ &  $\cdots$ & 815.4969 & 7.0925 &  814.9610 & 7.7355 & 814.4258 & 8.0823 & 814.9361 & 4.3494 \\
$\cdots$ &  $\cdots$ & 817.8851  & 3.0128 & $\cdots$ &  $\cdots$ &  819.2914 & 2.7160 &  $\cdots$ & $\cdots$ & $\cdots$ & $\cdots$ \\
$\cdots$ & $\cdots$ & 828.4669  & 3.8807  &  825.1054 & 4.6958 & 825.3679 & 2.9158 &  $\cdots$ & $\cdots$ & 825.6349 & 2.5266 \\
$\cdots$ & $\cdots$ & $\cdots$ & $\cdots$ &  829.9834 & 5.2491 & 830.8310 & 10.9648 & 829.1089 & 12.0168 & 830.4214 & 4.3077 \\
849.7289  & 6.6439 & $\cdots$ & $\cdots$ & $\cdots$ & $\cdots$ & $\cdots$ & $\cdots$  &  $\cdots$ & $\cdots$ & $\cdots$ & $\cdots$\\
862.2140  & 4.8798 & $\cdots$ & $\cdots$ & $\cdots$ &  $\cdots$ & $\cdots$ & $\cdots$ &  $\cdots$ & $\cdots$ & $\cdots$ & $\cdots$\\
875.2320  & 3.8219 & $\cdots$ & $\cdots$ & $\cdots$ &  $\cdots$ & $\cdots$ & $\cdots$ &  $\cdots$ & $\cdots$ & $\cdots$ & $\cdots$\\
\hline\hline
\end{tabular}
\end{table*}

\subsubsection{Rotational splittings}
\label{triplets}

White dwarf stars are considered slow rotators, with rotation periods that range from hours to days \citep{2017EPJWC.15201011K, 2017ApJS..232...23H}. Rotation leads to a breaking of the  degeneracy in pulsation frequencies, causing a single $\ell$ mode to separate into the $2\ell+1$ components in the azimuthal order $m$ \citep[e.g.][]{1989nos..book.....U}.

For slow rotation we can consider that the frequency separation between the central $m=0$ component and the $\pm m$ components are equal, and thus, a rotation period can be estimated following the equation \citep{1949ApJ...109..149C, 1951ApJ...114..373L}:
\begin{equation}
    \frac{1}{P_{\rm rot}} = \frac{\Delta \nu_{k,\ell, m}}{m (1-C_{k\ell})}
    \label{rotation}
\end{equation}
where $m$ is the azimuthal number and $C_{k\ell}$ is the rotational splitting coefficient given by:
\begin{equation}
    C_{k,\ell}=\frac{\int_0^{R_*} \rho r^2 [2\xi_r\xi_t + \xi_t^2] dr}{\int_0^{R_*} \rho r^2[\xi_r^2 + \ell(\ell + 1)\xi_t^2]dr}
\label{ckl}
\end{equation}
where $\rho$ is the density, $r$ is the radius and $\xi_r$ and $\xi_t$ are the radial and horizontal displacement of the material \citep[see][for details]{1989nos..book.....U}. In addition, the spherical degree and azimuthal order of the modes can be identified from the rotationally split multiplets present in the FT (e.g., \citealt{1991ApJ...378..326W,1994ApJ...430..839W}).  In the asymptotic regime, for high radial order modes, the value of the rotational splitting coefficient is $C_{k\ell} \sim 1/\ell(\ell+1)$, being $\sim 0.5$ and $\sim 0.166$, for $\ell=1$ and $\ell=2$ modes.

\begin{table}
\centering
\caption{Frequency, period, frequency separation and $\ell$ and $m$ identification for multiplets components found for PG 1541+651. The values in italic are computed from the observed frequencies. The values of the periods longer 800 sec are computed as a simple mean of the values for blocks 41, 47-48 and 50-51.}
\label{table-multiplets}
\begin{tabular}{ccccc}
\hline\hline 
Freq  & $\Pi$ & $\Delta\mu$ & $\ell$ & $m$\\
$\mu$Hz & s & $\mu$Hz & &\\
\hline
2493.744 & 401.004 & 6.609 & 1 & +1 \\
{\it 2487.135} & {\it 402.069} & - & 1 & 0 \\
2480.526 & 403.140 & 6.609 & 1 & -1 \\
\hline
1846.965 & 541.429 & 6.845 & 1 & +1 \\
{\it 1840.015} & {\it 543.474} & - & 1 & 0 \\
1833.065 & 545.534 & 7.036 & 1 & -1 \\
\hline
1471.462 & 679.596 & 6.580 & 1 & +1 \\
{\it 1464.882} & {\it 682.649} & - & 1 & 0 \\
1458.302 & 685.728 & 6.580 & 1 & -1 \\
\hline
1233.530 & 810.681 & 6.467 & 1 & -1 \\
1227.053 & 814.961 & - & 1 & 0 \\
1220.567 & 819.291 & 6.496 & 1 & -1 \\
\hline
1248.581 & 800.909 & 21.528 & ? & + \\
1227.053 & 814.961 & - & ? & 0 \\
1204.857 & 829.974 & 22.196 & ? & - \\
\hline
1176.846 & 849.729 & 17.041 & ? & +\\
1159.805 & 862.214 & - & ? & 0 \\
1142.554 & 875.232 & 17.251 & ? & -\\ 
\hline\hline
\end{tabular}
\end{table}

\begin{figure}
	\includegraphics[width=0.5\textwidth]{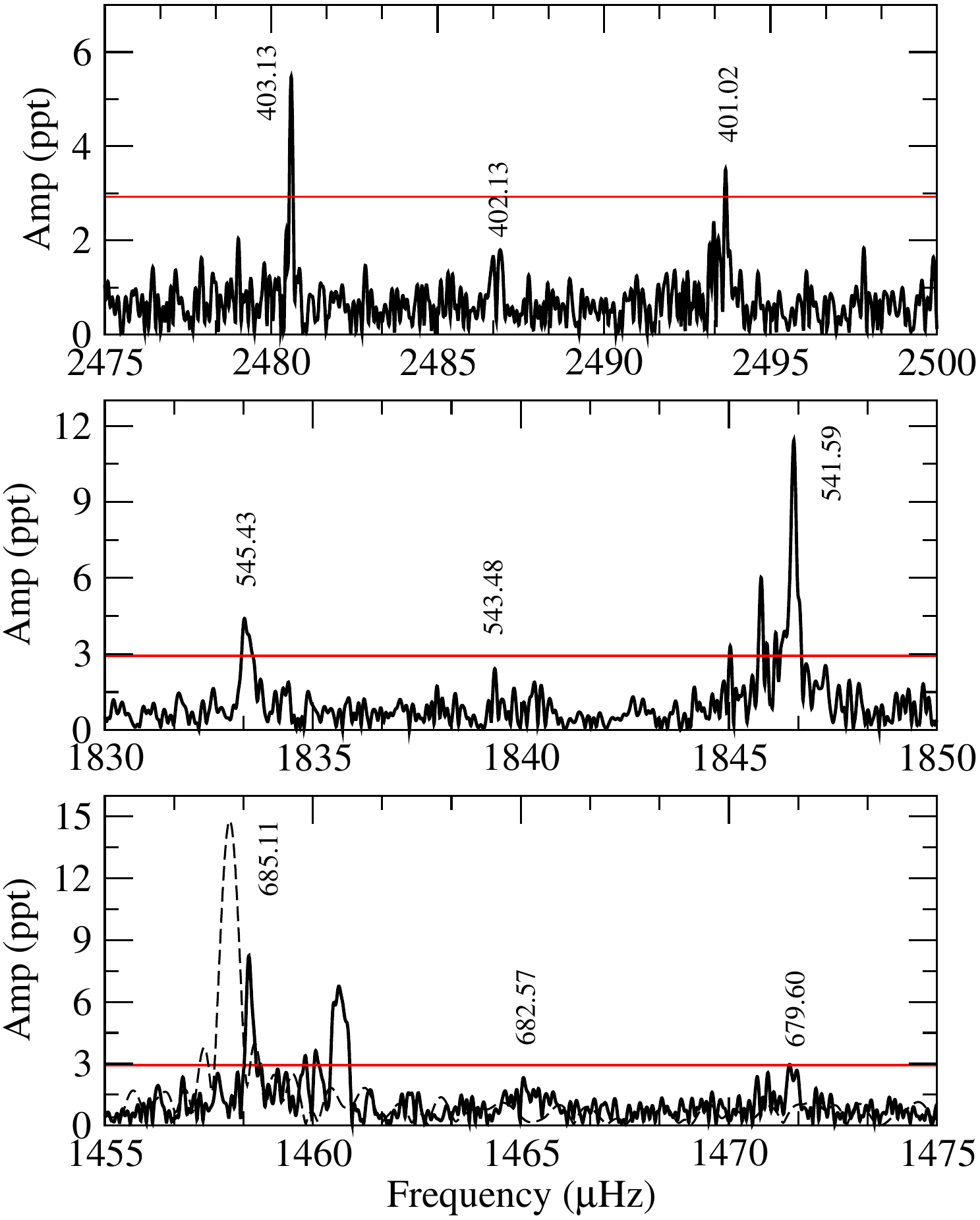}
    \caption{Portions of the Fourier transform for PG 1541+651 for the identified triplets with the shorter periods with central components at $\sim$ 402 s (top), $\sim$ 543 s (middle) and $\sim$ 683 s (bottom) for the observation from Sectors 21 to 24 and 120s--cadence. The red line correspond to the 5$\langle {\rm A}\rangle$ detection limit. For the bottom panel, we include the data for Sector 41 with 20s--cadence (dashed line) for completeness. The frequency separation between the central and prograde and retrograde components is between 6 and 7 $\mu$ Hz for all triplets shown in this figure. }
    \label{dipole}
\end{figure}

By analysing the frequency spectrum for PG~1541+651 from the TESS data, we identify seven possible multiplets. 
The components for each multiplet are listed in Table~\ref{table-multiplets}, along with the corresponding harmonic degree ($\ell$) and the azimuthal order ($m$). We identify four of them as triplets, with harmonic degree $\ell =1$, with a frequency separation between 6 and 7\,$\mu$Hz.
The regions of interest in the Fourier transform are shown in Figure~\ref{dipole} for the modes with central components at $\sim$ 2487\,$\mu$Hz (top), $\sim$1840\,$\mu$Hz (middle) and $\sim$1465\,$\mu$Hz (bottom) for the data from Sectors 21 to 24 and 120 s-cadence.

\begin{figure}
	\includegraphics[width=0.5\textwidth]{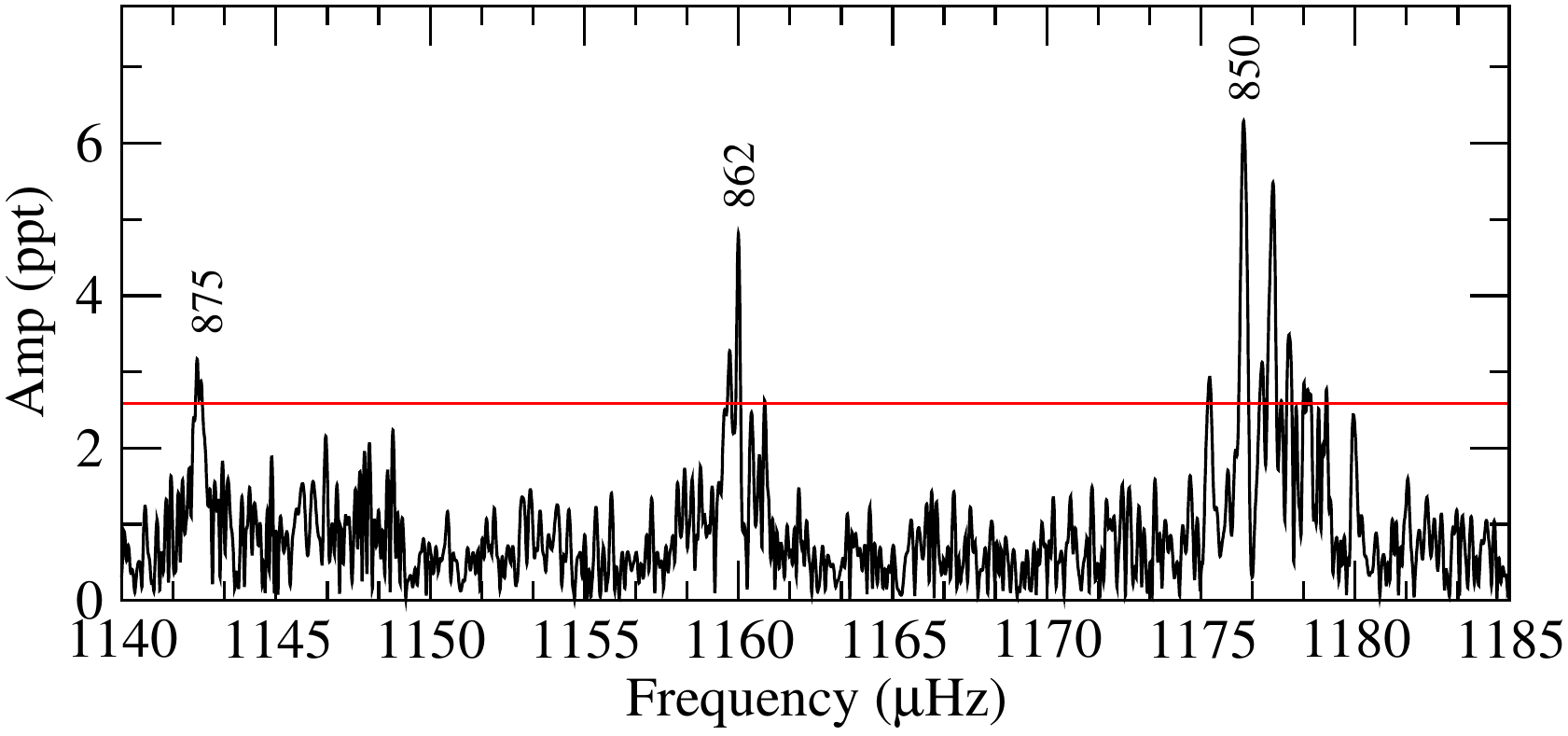}
    \caption{Fourier transform for PG 1541+651 for the region between 1140 and 1185 $\mu$Hz for PG 1541+651, for Sectors 14-17 and 120s--cadence. The red line correspond to the 5$\langle {\rm A}\rangle$ detection limit. }
    \label{long-dipole}
\end{figure}

The remaining two multiplets (last ones in Table~\ref{table-multiplets}) show a frequency separation between the components that is much larger than the others, and thus we do not assign a harmonic degree. The region of the FT corresponding to the multiplet with a central component at a frequency of 1159.81\,$\mu$Hz is depicted in Figure~\ref{long-dipole}. The frequency separation is $\sim$17\,$\mu$Hz, which is larger than two times the separation between the multiplets identified as dipole modes.

\begin{figure}
	\includegraphics[width=0.53\textwidth]{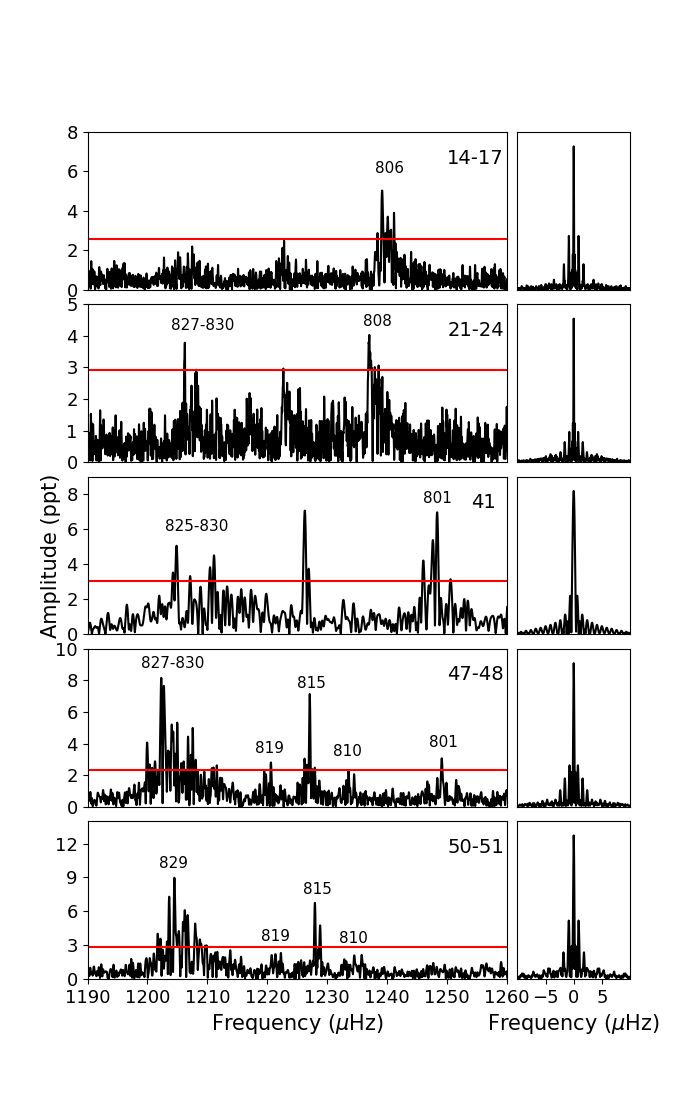}
	\caption{Fourier transform for the region between 1190 and 1260$\,\mu$Hz for PG~1541+651, for Sectors 14-17, 21-24, 41, 47-48 and 50-51 from top to bottom. Their respective spectral windows are shown on the right side. The red line correspond to the 5$\langle {\rm A}\rangle$ detection limit. The approximate values and ranges for the corresponding periods are depicted in the plot. On the right panel we depict the spectral window for each block.}
	\label{quadrupole}
	\end{figure}

As stated before, the region of the FT for frequencies around 1225\,$\mu$Hz shows a quite complex structure that changes from one block to the next. This can be seen from Figure~\ref{quadrupole}, where we show the region of interest for all blocks. From the first block, corresponding to Sectors 14 to 17, we only detect one peak at $\sim$806~s. As we move forward, other components seem to gain energy and reach amplitudes above the detection limit. In particular, components with higher frequencies (shorter periods) seem to lose energy as we move to more recent sectors, with the component centred at $\sim$829~s (1206 $\mu$Hz) being the one with the highest amplitude for the data corresponding to the 50-51 block. If we consider the combined peaks detected in the five blocks, we can identify five periods, as indicated in the fourth panel in Figure~\ref{quadrupole}, corresponding to the data from Sectors 47 and 48. We identify these collections of frequencies as two possible multiplets centred at $\sim$815~s (1226 $\mu$Hz). The first multiplet is identified as a $\ell =1$ mode, with a $\Delta \nu \sim$ 6.5$\,\mu$Hz (see Table~\ref{table-multiplets}). The second multiplet shows a separation of $\sim$22$\,\mu$Hz.

\subsection{BPM~31594}

\begin{figure*}
	\includegraphics[width=\textwidth]{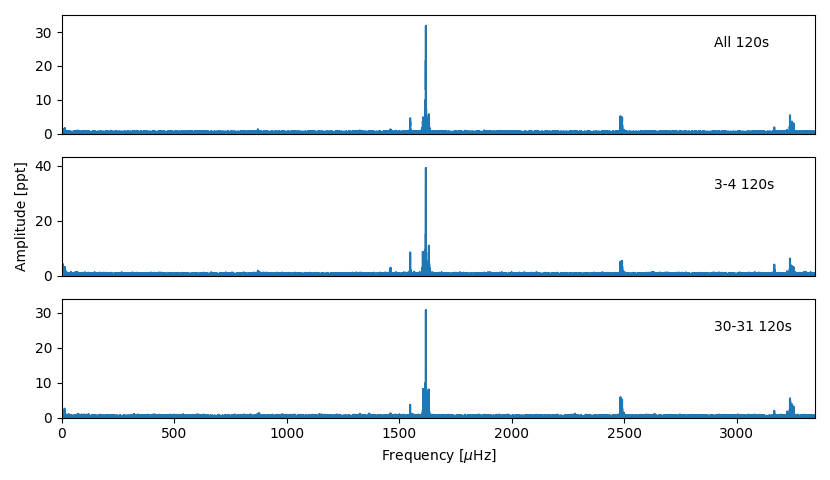}
    \caption{Fourier transform for the two blocks of data for BPM 31594. Top panel corresponds to the FT of the concatenated data. The middle panel corresponds to the data from Sectors 3 and 4 and 120s--cadence, while the bottom panel shows the FT for the data from Sectors 30 and 31 with  120s--cadence. Note that the amplitude scale is different for each plot. }
    \label{FT-BPM}
\end{figure*}

BPM~31594 was observed by TESS in Sectors 3 and 4 with 120~s cadence, and in Sectors 30 and 31 with 20~s-cadence. Thus, we separate the data into two blocks, corresponding to continuous observation runs. Figure \ref{FT-BPM} shows the FT for the two blocks, corresponding to Sectors 30 and 31 (bottom panel) and Sectors 3 and 4 (middle panel). The FT for the concatenated data is shown in the top panel of this figure.

\begin{figure}
	\includegraphics[width=0.5\textwidth]{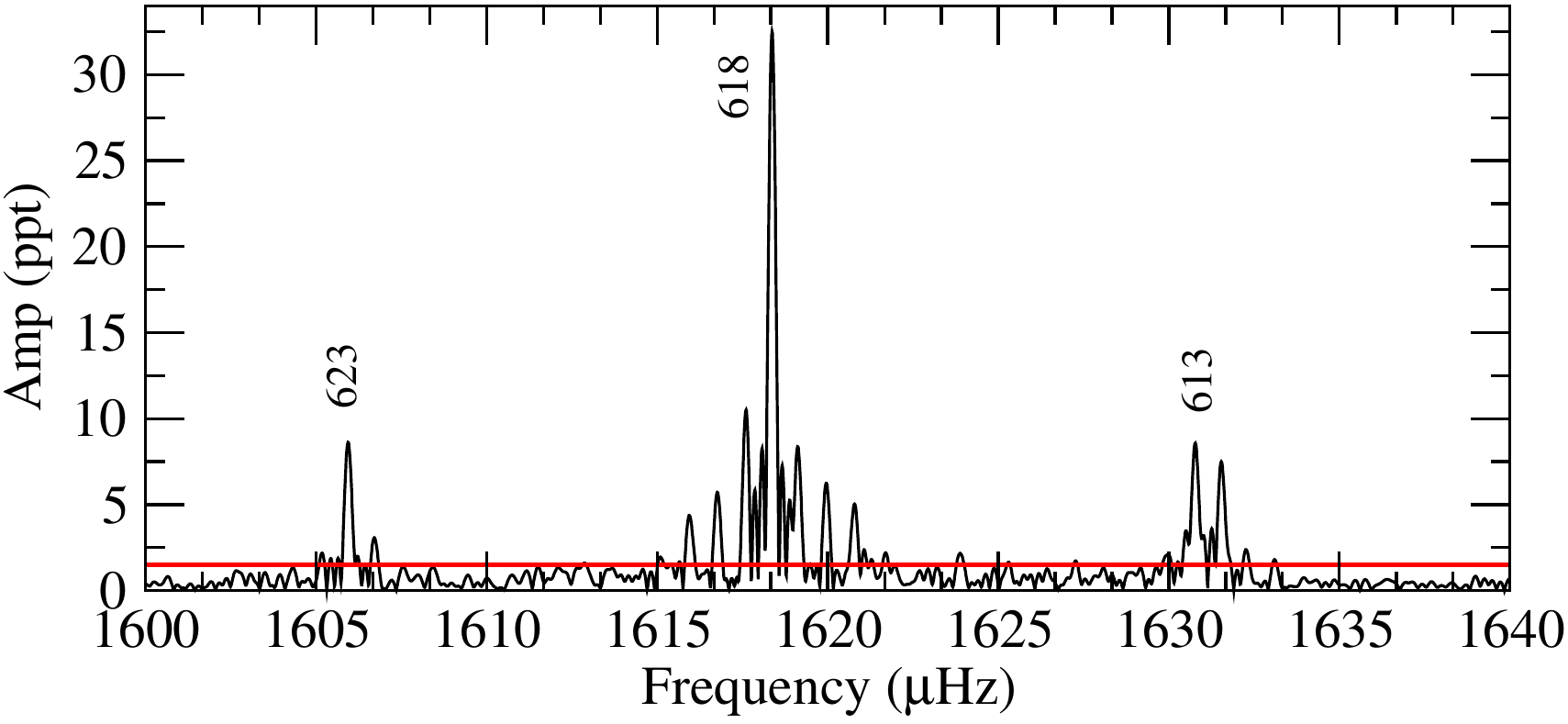}
    \caption{Fourier transform for BPM~31495 for the region between 1600 and 1620\,$\mu$Hz, for Sectors 30-31 at 20s--cadence. The red line correspond to the 5$\langle {\rm A}\rangle$ detection limit. }
    \label{BPM-dipole}
\end{figure}

The detected frequencies, periods and amplitudes for each block are listed in Table \ref{table-BPM-obs}. From the 03-04 block, we detect 6 periods, with a dominant peak at 617.89 s. From block 30-31, we detect five additional periods, and confirm the ones present in block 03-04. The mode with the highest amplitude is by far the one with a frequency of 1618.4 $\mu$Hz ($f_3$), in agreement with the results from \citet{1976ApJ...210L..35M} and \citet{1992MNRAS.258..415O}. We also detected the prograde and retrograde components for this frequency, shown in Figure \ref{BPM-dipole}, and thus we identify this mode as a dipole $\ell =1$ mode. Among the detected frequencies,  we look for linear combinations and harmonics by computing combinations of the detected frequencies, considering that their amplitude cannot be larger than that of the parent modes \citep{1995A&A...296..405B}. We identify two harmonics of the main period, and three linear combinations of the main peak and $f_2$, identified as such in the last column of Table \ref{table-BPM-obs}. Note that the peak corresponding to the frequency 3166.7 $\mu$Hz reported by \citep{1976ApJ...210L..35M} is present in the TESS data for both blocks.  On the other hand, sub-harmonics of the main period, previously reported by \citet{1992MNRAS.258..415O}, are not present in the TESS data.

\begin{table*}
\centering
\caption{List of frequencies, periods and amplitudes detected for BPM 31594.}
\label{table-BPM-obs}
\begin{tabular}{cccccccc}
\hline\hline
& 03-04 &   &     & 30-31  & & ID  \\ 
\hline
Freq & $\Pi$ & Amp & Freq & $\Pi$ & Amp & \\

$\mu$Hz & s & ppt & $\mu$Hz & s & ppt & \\
\hline
3166.8722 & 315.7670 & 4.0095 & 3166.6852 & 315.7876 & 2.3515  & $f_1$\\
2487.8989 & 401.9456 & 5.7188 & 2488.6026 & 401.8319 & 6.3873  & $f_2$\\
1631.8405 & 612.8050 & 10.9446 & 1630.7763 & 613.2049 & 7.9732 & $f_3^+$ \\
1618.3989 & 617.8946 & 39.114 & 1618.3725 & 617.9047 & 32.7478 & $f_3$\\
1604.9801 & 623.0607 & 8.8420 & 1605.9453 & 622.6862 & 8.6200 & $f_3^-$ \\
1548.4505 & 645.8069 & 8.4213 & 1548.3318 & 645.8564 & 3.6726  & $f_4$\\
1461.1146 & 684.8069 & 2.9050 & 1460.7312 & 684.5886 & 1.6364  & $f_5$\\
$\cdots$ & $\cdots$ & $\cdots$ & 877.2352 & 1139.9452 & 1.5551 & $f_6$\\
\hline
$\cdots$ & $\cdots$ & $\cdots$ & 5726.9279 & 174.6137 & 1.7852 & 2$f_3 + f_2$\\
$\cdots$ & $\cdots$ & $\cdots$ & 4858.3346 & 205.8318 & 2.9610 & 3$f_3$\\
$\cdots$ & $\cdots$ & $\cdots$ & 4108.3563 & 243.4063 & 2.6466 & $f_3 + f_2$ \\ 
3236.7978 & 308.9473 & 6.2662 &  3236.7464 & 308.9522 & 7.1622 & 2$f_3$ \\
$\cdots$ & $\cdots$ & $\cdots$ & 871.7971 & 1147.0558 & 1.5437 & $f_2 - f_3$ \\
\hline\hline
\end{tabular}
\end{table*}

\section{Asteroseismology}
\label{sec4}

In this section, we present a detailed asteroseismological study of PG~1541+651 and BPM~31594.
We employ a grid of DA white dwarf representative models, product of fully evolutionary computations that consider the evolution of the progenitor star from the main sequence to the cooling curve. The models were computed using the LPCODE evolutionary code \citep[see][for details]{2010ApJ...717..897A, 2010ApJ...717..183R,2013ApJ...779...58R, 2015MNRAS.450.3708R}. The model grid covers the mass region where C/O--core white dwarfs are expected, with stellar masses between 0.493 and 1.05\,$M_{\odot}$. The inner chemical profile is determined by the model evolution, thus the internal structure, from the C/O core to the hydrogen envelope, are consistent. In addition, models with different thickness of the hydrogen layers are included \citep{1996ApJ...468..350B,1949ApJ...109..149C}, with hydrogen masses ranging from $4\times 10^{-4} M_*$ to $\sim 10^{-10} M_*$, where the upper limit depends on the stellar mass, and is set by residual nuclear burning at the base of the hydrogen layer \citep{2012MNRAS.420.1462R, 2019MNRAS.484.2711R}. We do not consider hydrogen envelopes thinner than $10^{-10} M_*$ since, by the time the white dwarf reaches the instability strip, the outer convective zone will be deep enough to mix the hydrogen into the much ($\sim 100x$) more massive helium layer, turning the star into a DB white dwarf \citep{2013ApJ...779...58R, 2020MNRAS.492.5003O, 2020MNRAS.492.3540C}. Our current model grid has 616 white dwarf cooling sequences, corresponding to 35 stellar masses, totalling $\sim 74\, 000$ models. For each model within the instability strip, we computed adiabatic non-radial pulsations using the adiabatic version of the LP-PUL pulsation code \citep[see][for details]{2006A&A...454..863C}. This model grid has been largely used in the past to perform asteroseismological studies of DA pulsating white dwarfs \citep[][]{2012MNRAS.420.1462R, 2022MNRAS.511.1574R}.

To search for the asteroseismological model for each target, we seek for the theoretical model that minimizes a quality function, given by:
\begin{equation}
    \chi^2 = \frac{1}{N-1}\sqrt{\sum_{i=1}^{N} \left(\Pi_i^{\rm teo}- \Pi_i^{\rm obs}\right)^2}
    \label{chi2}
\end{equation}
where $N$ is the number of observed period and $\Pi_i^{\rm teo}$ is the theoretical period that better fits the observed period $\Pi_i^{\rm obs}$. We also compute other quality functions, always obtaining similar results \citep[see e.g.][]{1998ApJS..116..307B,2008MNRAS.385..430C,2009A&A...499..257C}. In our fit, we consider the identification of the harmonic degree, when present, and restrict our solutions using the stellar mass and effective temperature obtained from Gaia photometry and parallax.

\subsection{PG~1541+651}

The list of periods for PG~1541+651 considered for the asteroseismological fit is listed in Table~\ref{Periods-Pg-fit}. The period values are taken from Table~\ref{table-PG-obs}. For periods that are detected in more than one block, we consider the mean value, weighted by the amplitude in the FT. In case of multiplets, we consider the value of the period of the central component ($m=0$) from Table~\ref{table-multiplets}. For PG~1541+651 we identify 12 independent modes; four of them are identified positively as $\ell=1$ dipoles and one is identified as a $\ell=2$ quadrupole, given the number of detected components and the frequency separation between them (see section \ref{triplets}). 

\begin{table}
\centering
\caption{Periods list for PG~1541+651 used for asteroseismology. For the periods that appear as multiplets in the FT, we indicate the harmonic degree. Values for the modes marked in italic correspond to the central components of the multiplets shown in Figure \ref{quadrupole}.}
\label{Periods-Pg-fit}
\begin{tabular}{cccc}
\hline\hline 
ID & Freq & $\Pi$ & $\ell$ \\
   & $\mu$Hz & s & \\
\hline
$f_1$ & 3305.3641 & 302.5385 & - \\
$f_2$ & 2663.4520 & 375.7526 & - \\
$f_3$ & 2487.1350 & 402.0690 & 1 \\
$f_4$ & 1840.0153 & 543.4738 & 1\\
$f_5$ & 1758.6227 & 568.6268 & - \\
$f_6$ & 1608.2140 & 621.8078 & - \\
$f_7$ & 1464.8820 & 682.6488 & 1 \\
$f_8$ & 1368.9904 & 730.4653 & -\\
$f_9$ & 1279.9841 & 781.2597 & -\\
{\it $f_{10}$} & {\it 1227.0522} & {\it 814.9613} & {\it 1}\\
{\it $f_{11}$} & {\it 1227.0522} & {\it 814.9613} & {\it 2} \\
$f_{12}$ & 1159.8049 & 862.2140 & - \\
\hline\hline
\end{tabular}
\end{table}

As can be seen from Table~\ref{table-PG-obs}, most of the periods are not detected in all the data blocks. In fact, only four modes are detected in four of the five the blocks, whether it is the central components or at least one component of a multiplet. These modes are the ones identified by $f_3$, $f_4$, $f_7$  and $f_{10}$, with periods of 402.0690, 543.4732, 682.6488 and 814.9609~s. For the moment, we consider that the last period (814.96~s), corresponds to only one mode with an unknown harmonic degree. We first perform an asteroseismological fit using these four periods. The structural parameters of two models with the lowest value of $\chi ^2$ are listed in Table~\ref{PG-sis-1}. Both models are characterized by a stellar mass of 0.609\,$M_{\odot}$, but the hydrogen envelope mass is quite different. The model with the lowest value of $\chi^2$ shows a thin hydrogen envelope of $4.7\times 10^{-9} M_*$ and an effective temperature of $11\, 800$~K, closer to the blue edge of the instability strip for this stellar mass. The second model is characterized by a hydrogen envelope almost three orders of magnitude thicker and an effective temperature of $11\, 290$~K, closer to the red edge of the instability strip, which is in agreement with the period values larger than $\sim 400$~s \citep{2006ApJ...640..956M}. The large difference in the hydrogen content of the two seismological fits can be interpreted in terms of the core-envelope symmetry \citep{2003MNRAS.344..657M} and the differences in the chemical structures characterizing both models.

\begin{table*}
\centering
\caption{Best fitting models for PG~1541+651 considering the four recurring periods. The stellar mass, hydrogen envelope and effective temperature are listed in columns 1, 2 and 3, respectively. The theoretical periods are listed in column 4, along with the harmonic degree and radial order. The value of the quality function is listed in the last column.}
\label{PG-sis-1}
\begin{tabular}{ccccc}
\hline\hline 
M ($M_{\odot}$ & $\log(M_H/M_*)$ & $T_{\rm eff}$ [K] & $\Pi$ [s] $(\ell,k)$ & $\chi^2$ \\
\hline
0.609 & $-$8.33 & 11800 & 397.5016 (1,5), 540.5612 (1,8), 686.4946 (1,11), 813.7084 (2,25) & 2.2535 \\
0.609 & $-$5.24 & 11290 & 400.3013 (1,6), 550.4957 (1,9), 678.4833 (1,12), 814.4608 (1,15) & 2.7890  \\
\hline\hline
\end{tabular}
\end{table*}

\begin{figure}
	\includegraphics[width=0.5\textwidth]{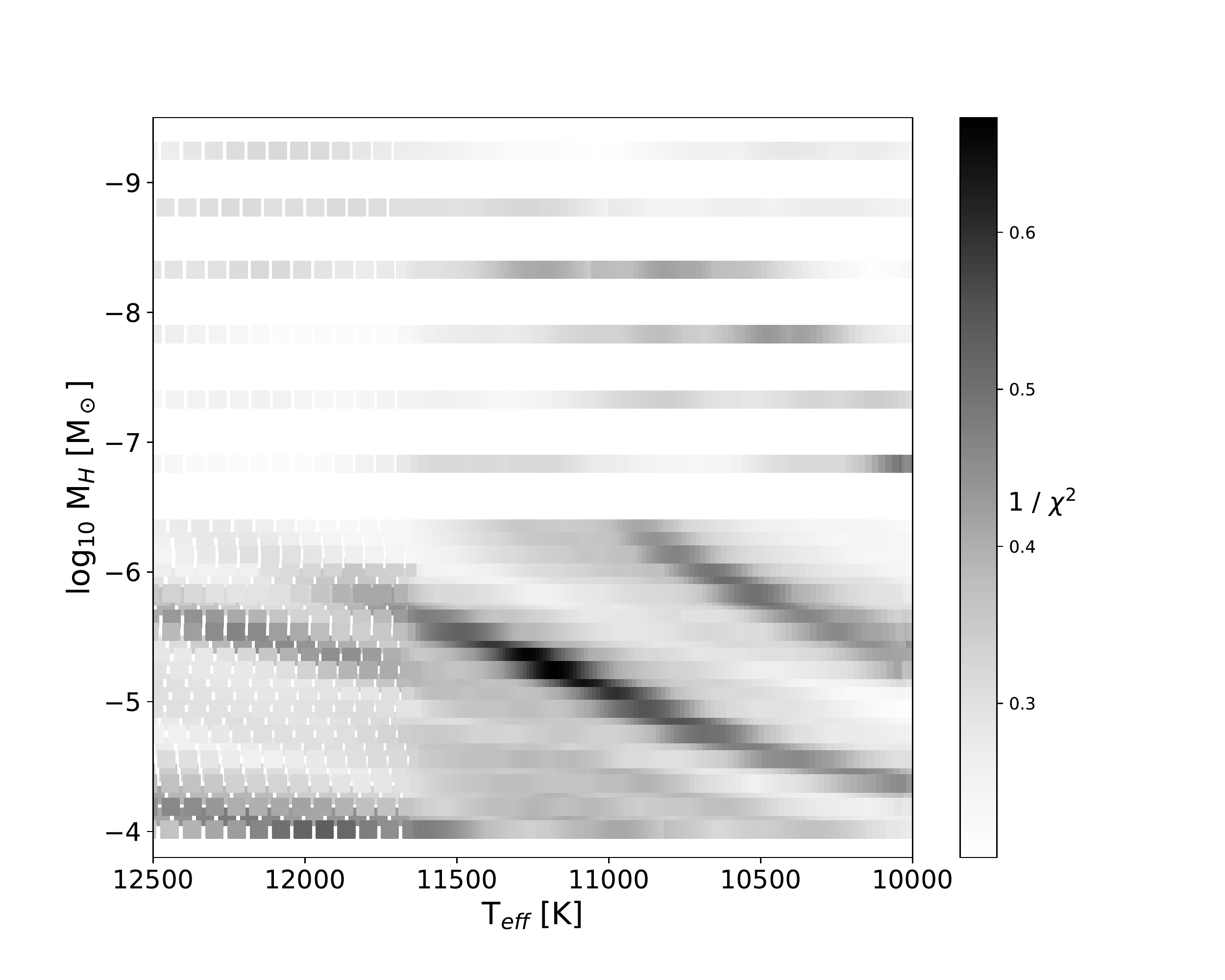}
    \caption{Contour map with the value of the inverse of the quality function (gray scale) for PG 1541+651, considering models with stellar mass of 0.609 M$_{\odot}$. The quality function is that given by equation \ref{chi2}, plotted as a function of the effective temperature and the logarithm of the hydrogen envelope mass.  }
    \label{PG-fit}
\end{figure}

\begin{figure}
	\includegraphics[width=0.45\textwidth]{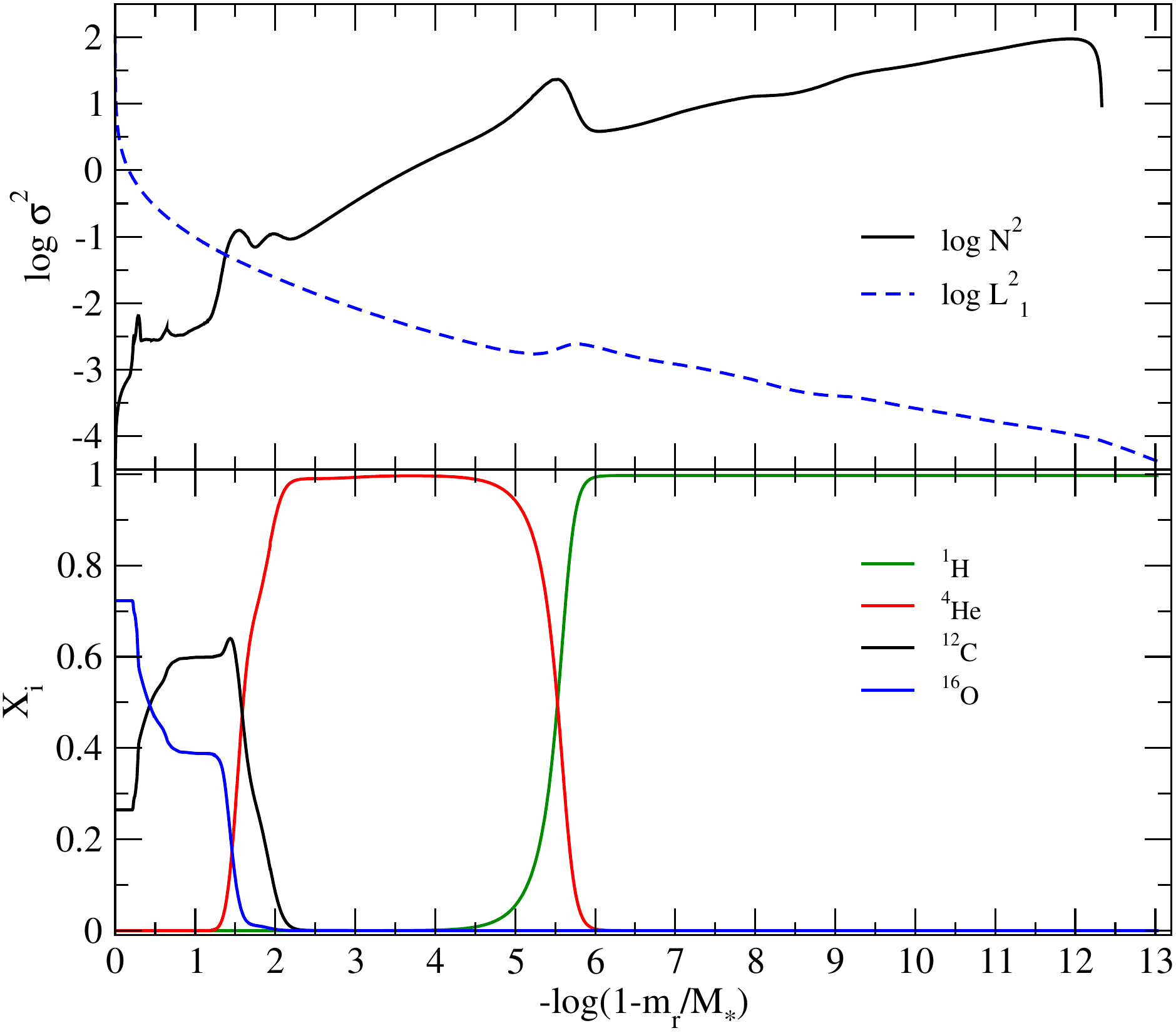}
    \caption{The square of the Brunt-V\"ais\"al\"a and the Lamb frequencies (top panel) and the chemical profiles (bottom panel) for the best fit model for PG 1541+651 listed in Table \ref{PG-sis-2}, with a stellar mass of 0.609 $M_{\odot}$, effective temperature of $11\, 240$ K, and a hydrogen mass of $4.5 \times 10^{-6} M_*$.   }
    \label{PG-profile}
\end{figure}

Next, we perform an asteroseismological fit considering all 12 periods detected for PG~1541+651, listed in Table~\ref{Periods-Pg-fit}. In this case, we consider that the period of $\sim 815$~s corresponds to the central component of two multiplets, that have harmonic degrees $\ell=1$ and $\ell=2$. The structural parameters characterizing the best fit model are listed in Table~\ref{PG-sis-2}, along with the theoretical periods and the value of the quality function. Figure \ref{PG-fit} shows the inverse of $\chi^2$ as a function of the effective temperature and the mass of the hydrogen envelope, for sequences characterized by a stellar mass of the best fit model, being 0.609~M$_{\odot}$. The effective temperature for the best fit model (minimum of $\chi^2$) is in agreement with the determination from \citet{2021MNRAS.508.3877G}, within the uncertainties. 
The chemical profile and the run of the Brunt--V\"ais\"al\"a and the Lamb frequencies for the best fit model are shown in Figure~\ref{PG-profile}.

\begin{table}
\centering
\caption{Best fitting model for PG~1541+651 considering all detected periods. The stellar mass, effective temperature and hydrogen envelope are listed in column 2. The theoretical periods, harmonic degree and radial order are listed in columns 3, 4 and 5, respectively. The value of the rotation kernel and the rotation period are listed in the columns 6 and 7, respectively, for the modes identified as multiplets. The value of the quality function is listed in the last row of column 2.}
\label{PG-sis-2} 
\begin{tabular}{cccccccc}
\hline\hline
Model & & $\Pi$ [s] & $\ell$ & $k$ &  $C_{k \ell}$  & P$_{\rm rot}$ [h]\\
\hline 
M ($M_{\odot}$)   & 0.609   & 305.6120 & 2 & 8 & & \\
$T_{\rm eff}$ [K] & 11240   & 375.9419 & 2 & 11 & & \\
$\log(M_H/M_*)$  & $-$5.35  & 402.9660 & 1 & 6 & 0.4877 & 21.53\\
  $\chi^2$       & 1.5037   & 552.9017 & 1 & 9 & 0.4521 & 21.93\\
                 &          & 564.3766 & 2 & 18 &  &\\  
                 &          & 616.4197 & 2 & 20 & &\\
                 &          & 680.4286 & 1 & 12 & 0.4587 & 22.85 \\
                 &          & 726.3037 & 2 & 24 & &\\
                 &          & 782.4958 & 2 & 26 & &\\
                 &          & 809.2072 & 2 & 27 &  0.1626 & 21.28* \\
                 &          & 821.7833 & 1 & 15 &  0.4946 & 21.66 \\
                 &          & 867.1427 & 2 & 29 & 0.1642  & 27.08* \\
\hline\hline
\end{tabular}
\end{table}

\begin{figure}
	\includegraphics[width=0.45\textwidth]{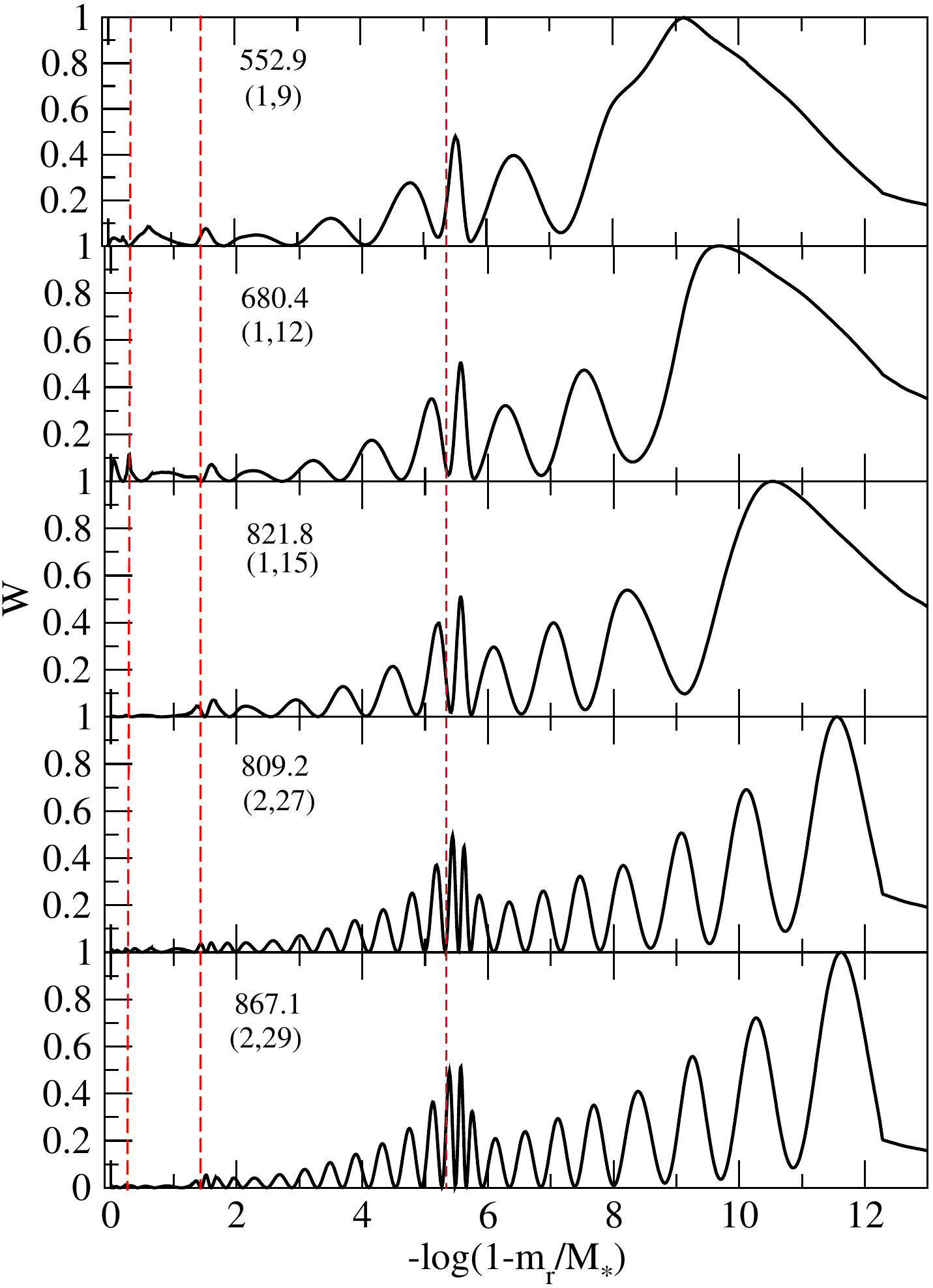}
    \caption{Weight function $W$ corresponding to the modes with $(\ell,k)$ - (1,9), (1,12), (1,15), (2,27) and (2,29) of the asteroseismic model for PG~1541+651. Vertical dashed lines correspond to the position of the main chemical transitions (see Fig. \ref{PG-profile}).  }
    \label{PG-W}
\end{figure}

Figure \ref{PG-W} shows the run of the weight function $W$ for five of the six modes identified as central components of multiplets. The vertical red lines correspond to the position of the each chemical transition, being, from right to left, the H--to--He transition, the base of the He buffer, and the point where the carbon abundance is larger than the oxygen abundance in the C/O core (see Fig. \ref{PG-profile} for details). Note that the weight function is very sensitive to the position of the H-to-He transition for all modes depicted in Figure \ref{PG-W}.

Finally, each theoretical mode has an associated value of the rotational splitting coefficient $C_{k, \ell}$. The values for the  $C_{k, \ell}$ for the modes identified as multiplets are listed in column 6 of Table \ref{PG-sis-2}. Combining this value with the observed frequency separation for each mode (third column of Table \ref{table-multiplets}), we can estimate the rotation period using equation \ref{rotation}. The rotation periods are listed in the last column of Table \ref{table-multiplets}. For the modes identified as dipoles, with $\ell =1$, the values for the rotation period are in agreement with each other, giving a value of $\sim$ 22 h. The rotation period obtained for the mode with theoretical period 821.7833 s, identified as $\ell=2$, shows a similar value, if we consider that the observed components correspond to the azimuthal orders $m=\pm 2$. Thus, if we consider the five multiplets with the shortest periods, we can estimate a rotation period of $\sim$ 22 h.  

The mode with a central component at 862.2140~s was also identified as a multiplet. The frequency separation in this case is $\Delta \nu_{\ell, m}\sim 17\,\mu$Hz. The best fit model for PG 1541+651 fits this mode as a quadrupole theoretical mode with $\ell=2$. The rotation period obtained for this mode gives $\sim 13.5$ h if we consider the components to be $m=\pm 1$, or 27.1~h if we consider the components to be $m=\pm 2$. 

\subsection{BPM~31594}

For BPM~31594 we detected nine frequencies in the FT, with six modes identified as independent modes. For our asteroseismological fit we consider the period values from the 30-31 block, and fix the harmonic degree of the main mode, with a period of 617.9~s, as $\ell=1$. The final list is shown in Table~\ref{periods-BPM-fit}.

\begin{table}
\centering
\caption{Periods list for BPM~31594 used for asteroseismology. For the period that appears as a multiplet in the FT, we indicate the harmonic degree.}
\label{periods-BPM-fit}
\begin{tabular}{cccc}
\hline\hline
ID & Freq & $\Pi$ & $\ell$ \\
   & $\mu$Hz & s & \\
\hline
$f_1$ & 3166.6852 & 315.7876 &  -\\
$f_2$ & 2488.6026 & 401.8319 &  - \\
$f_3$ & 1618.3725 & 617.9047 &  1 \\
$f_4$ & 1548.3318 & 645.8564 &  - \\
$f_5$ & 1460.7312 & 684.5886 & - \\
$f_6$ &  877.2352 & 1139.9452 & -  \\
\hline\hline
\end{tabular}
\end{table}

\begin{figure}
	\includegraphics[width=0.5\textwidth]{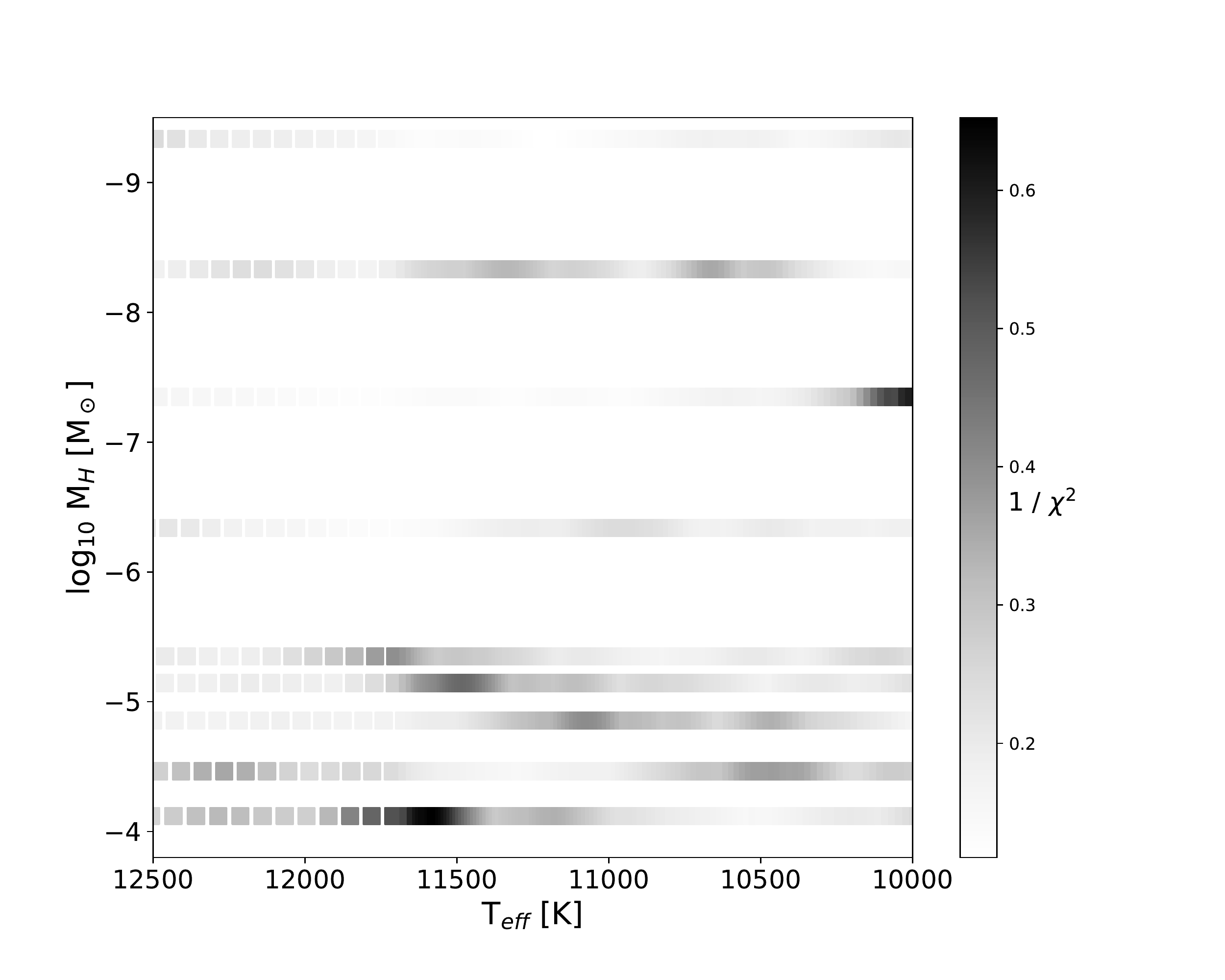}
    \caption{Contour map with the value of the inverse of the quality function (gray scale) for BPM 31594, considering models with stellar mass of 0.632 M$_{\odot}$. The quality function is that given by equation \ref{chi2}, plotted as a function of the effective temperature and the logarithm of the hydrogen envelope mass.  }
    \label{BPM-fit}
\end{figure}

\begin{figure}
	\includegraphics[width=0.45\textwidth]{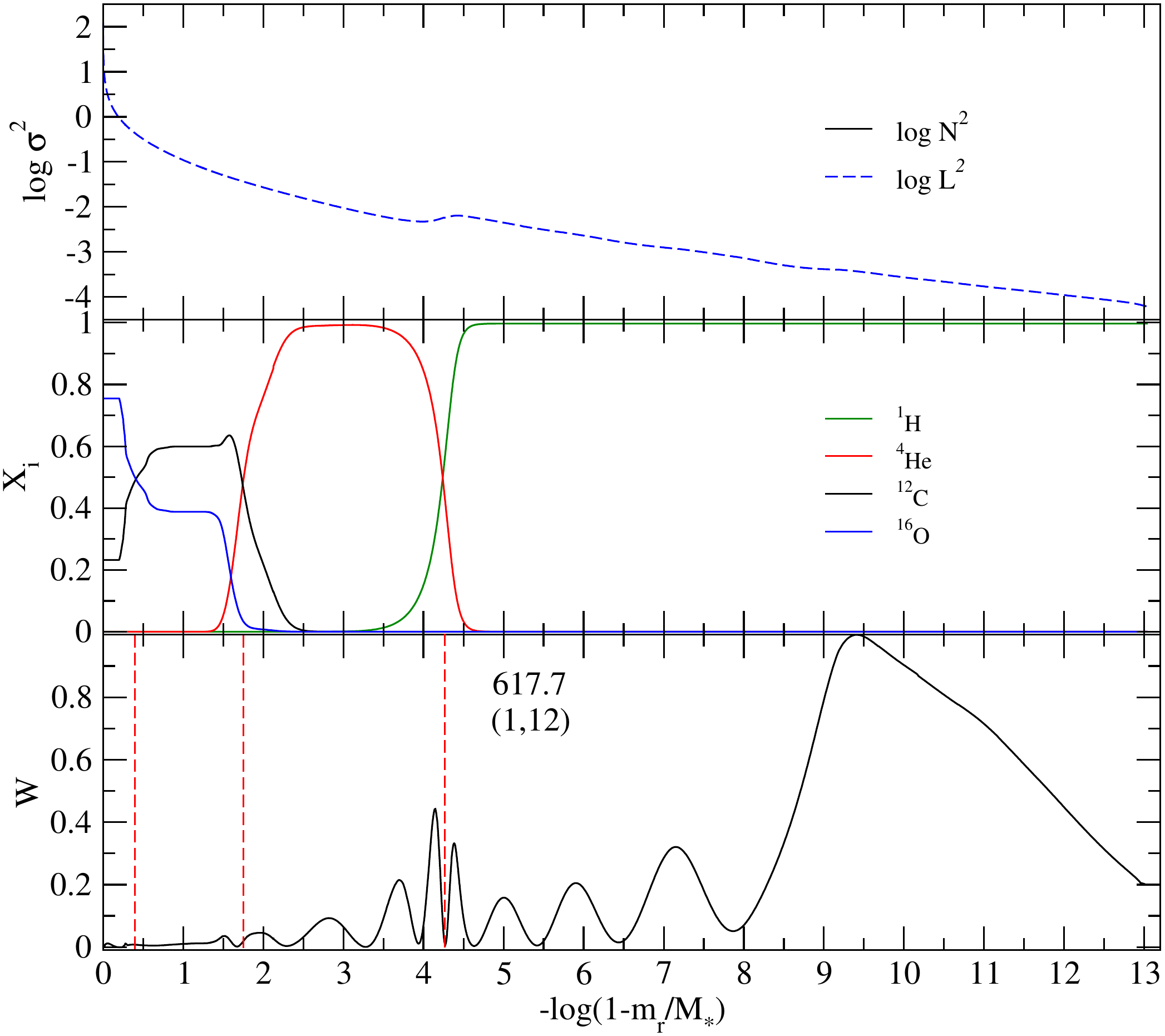}
    \caption{The square of the Brunt-V\"ais\"al\"a and the Lamb frequencies (top panel) and the chemical profiles (middle panel) for the best fit model for BPM~31594 listed in Table \ref{BPM-sis}, with a stellar mass of 0.632~$M_{\odot}$, effective temperature of $11\, 560$ K, and a hydrogen mass of $7.6 \times 10^{-5} M_*$. Run of the weight function $W$ (bottom panel) for the main mode, with a period of 617.7 s.  }
    \label{BPM-profile}
\end{figure}

From our asteroseismological fit, we found a minimum in the quality function for a stellar mass of 0.690\,$M_{\odot}$. However, the luminosity of the model leads to a distance of 37~pc, which is 7 pc closer than the distance determined with Gaia DR3 parallax, of 44.3~pc \citep{2021MNRAS.508.3877G}. We searched for a seismological solution that is compatible with the distance determination. The structure parameters and the theoretical periods of this model are listed in Table~\ref{BPM-sis}. Figure \ref{BPM-fit} depicts the inverse of $\chi^2$ as a function of the effective temperature and the mass of the hydrogen envelope, for sequences with a stellar mass of 0.632~M$_{\odot}$, where a solution family can be seen for the thickest envelope value. Figure \ref{BPM-profile} shows the run of the Brunt-V\"ais\"al\"a and Lamb frequencies (top panel) and the chemical profile for the best fit model (middle panel). In the bottom panel of Figure \ref{BPM-profile} we depict the weight function $W$ for the theoretical mode with a period of 617.7 s. This is the main observed mode, which is also identified as a triplet. Note that this mode is quite sensitive to the H/He transition, and thus to the mass of the hydrogen envelope.  

\begin{table}
\centering
\caption{Best fitting model for BPM~31594 considering all detected periods. The stellar mass, effective temperature and hydrogen envelope are listed in column 2. The theoretical periods, harmonic degree and radial order are listed in columns 3, 4 and 5, respectively. The value of the quality function is listed in the last row of column 2. The value of the rotation kernel is listed in the last column.}
\label{BPM-sis} 
\begin{tabular}{cccccc}
\hline\hline
$\#$ & & $\Pi$ [s] & $\ell$ & $k$ & $C_{k \ell}$ \\
\hline 
M ($M_{\odot}$)   & 0.632   & 315.2605 & 2 & 10 & \\
$T_{\rm eff}$ [K] & 11560   & 406.3396 & 1 & 7 & \\
$\log(M_H/M_*)$  & $-$4.12    & 617.7196 & 1 & 12 & 0.4825\\
  $\chi^2$        & 1.5500  & 644.1514 & 1 & 13 & \\
            &       & 679.0426 & 1 & 14  &  \\ 
            &       & 1142.0466 & 2 & 44 & \\
\hline\hline
\end{tabular}
\end{table}

From the asteroseismological fit we also get the value of the rotational splitting coefficient $C_{k \ell}$ (eq. \ref{ckl}) for all the theoretical periods. In particular, for the main mode fitted with a theoretical period of  617.7~s the value is $C_{1,12} = 0.4825$. Following equation \ref{rotation}, with $\Delta \nu_{k, \ell, m}$ = 12.4$\,\mu$Hz, we obtain a rotation period of 11.57~h.

\section{Conclusions}\label{conclusion}

In this work we present a detailed astroseismological study of two warm--like ZZ ceti stars, PG~1541+651 and BPM~31594, based on the photometric data obtained by the TESS mission. 

PG~1541+651 was observed in eight sectors with 120 s cadence, and also in five sectors with both 120 and 20 s cadence, from Sectors 14 to 51, with 20~s--cadence data for five sectors. We found 12 modes, four of them identified as triplets due to the presence of the rotational splitting components.  
From our asteroseismological fit we found a representative model characterized by a stellar mass of 0.609$M_{\odot}$, effective temperature of $11\, 240$ K and a mass of the hydrogen envelope of 4.5$\times 10^{-6}M_*$ (or 2.7$\times 10^{-6}M_{\odot}$). Due to the large number of observed modes we were able to break the degeneracy in $M_H$ in our asteroseismological fit.

BPM~31594 was observed during the first year of the TESS mission, in Sectors 3 and 4, and during the third year in Sectors 30 and 31. We identify six periods, being the dominant period as triplet. For this object we used the restriction in stellar mass given by the distance and found a best fit model characterized by $M_*=0.632 M_{\odot}$, $T_{\rm eff} = 11\, 560$ K and a $M_H = 7.6 \times 10^{-5}M_*$ (or 4.8$\times 10^{-5}M_{\odot}$), being the thickest hydrogen envelope model for this stellar mass.

Rotation periods for both objects were obtained from the frequency separation of the detected multiplets, being $\sim$ 22 h for PG 1541+651 and 11.57 h for BPM~31594, in agreement with the values reported for other ZZ Ceti stars.


\section*{Acknowledgements}

This study was financed in part by the Coordena\c{c}\~ao de Aperfei\c{c}oamento de Pessoal de N\'{\i}vel Superior - Brasil (CAPES) - Finance Code 001, Conselho Nacional de Desenvolvimento Cient\'{\i}fico e Tecnol\'ogico - Brasil (CNPq), and Funda\c{c}\~ao de Amparo \`a Pesquisa do Rio Grande do Sul (FAPERGS) - Brasil. KJB is supported by the National Science Foundation under Award AST-1903828. Financial support from the National Science Centre under project No.\,UMO-2017/26/E/ST9/00703 is acknowledged. J.J.H. acknowledges salary and travel support through {\em TESS} Guest Investigator Programs 80NSSC19K0378 and 80NSSC20K0592.
M.U. acknowledges financial support from CONICYT Doctorado Nacional in the form of grant number No: 21190886 and ESO studentship program. This paper includes data collected with the TESS mission, obtained from the MAST data archive at the Space Telescope Science Institute (STScI). Funding for the TESS mission is provided by the NASA Explorer Program. This work has made use of data from the European Space Agency (ESA) mission Gaia (\url{https://www.cosmos.esa.int/gaia}), processed by the Gaia Data Processing and Analysis Consortium (DPAC, \url{https://www.cosmos.esa.int/web/gaia/dpac/consortium}). Funding for the DPAC has been provided by national institutions, in particular the institutions participating in the Gaia Multilateral Agreement. 
This research has made use of NASA's Astrophysics Data System Bibliographic Services, and the
SIMBAD database, operated at CDS, Strasbourg, France.

\section*{Data Availability}

Data from TESS is available at the MAST archive \url{https://mast.stsci.edu/search/hst/ui/$#$/}. Ground based data will be shared on reasonable request to the corresponding author.

\bibliographystyle{mnras}
\bibliography{main} 

\end{document}